\documentclass[a4paper,12pt]{article}
\pdfoutput=1
\usepackage{feynmp-auto,expdlist}
\usepackage{amsmath, amsfonts, amssymb,mathbbol}
\usepackage{graphicx}
\usepackage{enumerate,cancel}
\usepackage{hyperref}
\usepackage{latexsym}
\usepackage{dsfont}
\usepackage{hepnicenames}
\usepackage{enumerate}
\usepackage{soul}
\usepackage[normalem]{ulem}
\usepackage{comment}
\usepackage{units}

\newcommand{\gfootnote}[1]{\footnote{\tiny\color{gray}#1}}
\renewcommand{\gfootnote}[1]{}
\usepackage{soul}

\usepackage{mathrsfs,graphicx,rotating,amsmath,amsfonts,mathtools,booktabs,amssymb,wasysym}
\usepackage{hyperref}\usepackage{slashed}
\usepackage[nosort]{cite}
\usepackage[table,xcdraw,dvipsnames]{xcolor}
\usepackage{bm}
\usepackage{graphicx}
\usepackage{multirow,multicol}
\hypersetup{colorlinks,bookmarksopen,bookmarksnumbered,
	linkcolor=blus,pdfstartview=FitH,urlcolor=rossos,citecolor=verde}
\allowdisplaybreaks

\renewcommand{\[}{\left[}

\newcommand{\Lag}{\mathscr{L}}

\newcommand{\mio}[1]{}

\newcommand{\med}[1]{\langle #1\rangle}

\newcommand{\bpm}{\begin{pmatrix}}
\newcommand{\epm}{\end{pmatrix}}

\usepackage{mathrsfs}

\newcommand{\fig}[1]{~\ref{fig:#1}}
\newcommand{\sfrac}[2]{#1/#2}

\newcommand{\beqa}{\begin{equation}\begin{aligned}}
\newcommand{\eeqa}{\end{aligned}\end{equation}}

\renewcommand{\Im}{{\rm Im}\,}
\renewcommand{\Re}{{\rm Re}\,}
\allowdisplaybreaks
\usepackage{multicol}
\usepackage{color}
\definecolor{rosso}{cmyk}{0,1,1,0.4}
\definecolor{rossos}{cmyk}{0,1,1,0.55}
\definecolor{rossoc}{cmyk}{0,1,1,0.2}
\definecolor{blu}{cmyk}{1,1,0,0.3}
\definecolor{blus}{cmyk}{1,1,0,0.6}
\definecolor{bluc}{cmyk}{1,1,0,0.1}
\definecolor{verde}{cmyk}{0.92,0,0.59,0.25}
\definecolor{verdec}{cmyk}{0.92,0,0.59,0.15}
\definecolor{verdes}{cmyk}{0.92,0,0.59,0.4}

\newcommand{\bp}{\bar{M}_{\rm Pl}}
\newcommand{\eq}[1]{~{\rm (\ref{eq:#1})}}

\newcommand{\GeV}{\,{\rm GeV}}

\newcommand{\Tr}{\,{\rm Tr}}

\newcommand{\beq}{\begin{equation}}
\newcommand{\eeq}{\end{equation}}

\newcommand{\bea}{\begin{eqnarray}}
\newcommand{\eea}{\end{eqnarray}}
\newcommand{\be}{\begin{equation}}
\newcommand{\ee}{\end{equation}}
\font\tenrsfs=rsfs10 at 12pt
\font\sevenrsfs=rsfs7
\font\fiversfs=rsfs5
\newfam\rsfsfam
\textfont\rsfsfam=\tenrsfs
\scriptfont\rsfsfam=\sevenrsfs
\scriptscriptfont\rsfsfam=\fiversfs
\def\be#1\ee{\begin{equation}#1\end{equation}}
\def\bl#1\el{\begin{align}#1\end{align}}
\def\ba#1\ea{\begin{align*}#1\end{align*}}

\renewenvironment{thebibliography}[1]
{\begin{multicols}{2}[\section*{\refname}]%
		\@mkboth{\MakeUppercase\refname}{\MakeUppercase\refname}%
		\list{\@biblabel{\@arabic\c@enumiv}}%
		{\settowidth\labelwidth{\@biblabel{#1}}%
			\leftmargin\labelwidth
			\advance\leftmargin\labelsep
			\@openbib@code
			\usecounter{enumiv}%
			\let\p@enumiv\@empty
			\renewcommand\theenumiv{\@arabic\c@enumiv}}%
		\sloppy
		\clubpenalty4000
		\@clubpenalty \clubpenalty
		\widowpenalty4000%
		\sfcode`\.\@m}
	{\def\@noitemerr
		{\@latex@warning{Empty `thebibliography' environment}}%
		\endlist\end{multicols}}

\newcommand{\SU}{\,{\rm SU}}
\newcommand{\SL}{\,{\rm SL}}

\newcommand{\U}{\,{\rm U}}

\makeatletter
\font\ital=cmu10

\def\hhref#1{\href{http://arxiv.org/abs/#1}{arXiv:#1}}
\usepackage{xstring}
\newcommand{\hhrefq}[1]{\IfSubStr{#1}{:}{\href{http://inspirehep.net/search?ln=en&ln=en&p=#1&of=hb&action_search=Search&sf=&so=d&rm=&rg=25&sc=0}{InSpire:#1}}{\hhref{#1}}}

\def\art{\@ifnextchar[{\eart}{\oart}}
\def\eart[#1]#2#3#4#5#6{{\rm #2}, {\em #3 \bf #4} {\rm (#6) #5} ({\em #1})}
\def\article{\@ifnextchar[{\earticle}{\oarticle}}
\def\oarticle#1#2#3#4#5#6{{\rm #1}, {\ital `#6'}, {\rm #2 #3 (#5) #4}}
\def\earticle[#1]#2#3#4#5#6#7{{\rm #2}, {\ital `#7'}, {\rm #3 #4 (#6) #5}  [\hhrefq{#1}]}
\def\hepart[#1]#2{{\rm #2, \sl#1}}
\def\heparticle[#1]#2#3{#2, {\ital `#3'} [\hhrefq{#1}]}
\newcommand{\doi}[1]{\href{http://dx.doi.org/#1}{[link]}}

\newcommand{\hhrefqq}[1]{\IfBeginWith{#1}{10.}{\href{https://doi.org/#1}{doi:#1}}{\hhrefq{#1}}}

\renewenvironment{thebibliography}[1]
{\begin{multicols}{2}[\section*{\refname}]%
		\@mkboth{\MakeUppercase\refname}{\MakeUppercase\refname}%
		\list{\@biblabel{\@arabic\c@enumiv}}%
		{\settowidth\labelwidth{\@biblabel{#1}}%
			\leftmargin\labelwidth
			\advance\leftmargin\labelsep
			\@openbib@code
			\usecounter{enumiv}%
			\let\p@enumiv\@empty
			\renewcommand\theenumiv{\@arabic\c@enumiv}}%
		\sloppy
		\clubpenalty4000
		\@clubpenalty \clubpenalty
		\widowpenalty4000%
		\sfcode`\.\@m}
	{\renewcommand{\@noitemerr}
		{\@latex@warning{Empty `thebibliography' environment}}%
		\endlist\end{multicols}}

\newcommand{\eqnsystem}[1]{
	\renewcommand{\@eqnnum}{{\rm (\thealphaequation)}}
	\renewcommand{\@@eqncr}{\let\@tempa\relax \ifcase\@eqcnt \def\@tempa{& & &} \or
		\newcommand{\@tempa}{& &}\or \newcommand{\@tempa}{&}\fi\@tempa
		\if@eqnsw\@eqnnum\refstepcounter{alphaequation}\fi
		\global\@eqnswtrue\global\@eqcnt=0\cr}
	\refstepcounter{equation} \let\@currentlabel\theequation \def\@tempb{#1}
	\ifx\@tempb\empty\else\label{#1}\fi
	\refstepcounter{alphaequation}
	\let\@currentlabel\thealphaequation
	\global\@eqnswtrue\global\@eqcnt=0 \tabskip\@centering\let\\=\@eqncr
	$$\halign to \displaywidth\bgroup \@eqnsel\hskip\@centering
	$\displaystyle\tabskip\z@{##}$&\global\@eqcnt\@ne
	\hskip2\arraycolsep\hfil${##}$\hfil& \global\@eqcnt\tw@\hskip2\arraycolsep
	$\displaystyle\tabskip\z@{##}$\hfil
	\tabskip\@centering&\llap{##}\tabskip\z@\cr}
\def\endeqnsystem{\@@eqncr\egroup$$\global\@ignoretrue} \makeatother

\newcounter{alphaequation}[equation]
\renewcommand{\thealphaequation}{\theequation\hbox to
	0.6em{\hfil\alph{alphaequation}\hfil}}

\definecolor{Gray}{gray}{0.95}

\begin{document}

\begin{center}
\bigskip\bigskip
{\bf\LARGE\color{red!60!black} Cosmological collider signals of \\[1ex] modular spontaneous CP breaking}\\
\bigskip\bigskip
{\bf Shuntaro Aoki}$^a$ and {\bf Alessandro Strumia}$^b$  \\[2ex]
{\it $^a$ RIKEN Center for Interdisciplinary Theoretical and Mathematical Sciences, Saitama, Japan\\[1ex]
$^b$ Dipartimento di Fisica, Universit\`a di Pisa, Italia}\\
\bigskip
{\bf\large\color{blus} Abstract}
\begin{quote}\large
We consider a modular-invariant extension of the Standard Model.
Assuming that the modulus is the inflaton, 
the CP-violating phases of the Yukawa couplings evolve during inflation.
This dynamics favours a Higgs condensate, so that
Standard Model fermions mediate a one-loop cosmological collider signal enhanced by chemical potentials.
The effect appears too small for next-generation experiments. 
We provide precise expressions for Dirac fermions with chemical potentials in de Sitter 
and compute the loop keeping the complete fermion modes.
\end{quote}
\end{center}
\thispagestyle{empty}

\tableofcontents

\section{Introduction}

In plausible extensions of the Standard Model, the observed violation of charge-conjugation times parity (CP) 
may arise from spontaneous symmetry breaking. 
In this framework, the fundamental theory is CP-invariant, and CP violation emerges dynamically when a scalar field, denoted by $\tau$, 
acquires an intrinsically complex vacuum expectation value. 

This picture is naturally realized in string constructions, where the underlying ten-dimensional theory is real. 
In such setups, CP violation in four dimensions can emerge from the geometry of compactification and is captured in the low-energy effective field theory by 
target-space modular invariance associated with the modulus $\tau$~\cite{Ferrara:1989bc,Dixon:1990pc,Ibanez:1992hc} (see~\cite{2501.16427} for a brief summary).
The supersymmetric version of such theories offers predictive flavour models~\cite{1706.08749} and
a novel solution to the strong CP problem~\cite{2305.08908}.
A novel baryogenesis mechanism arises if $\tau$ depends on time during the big bang~\cite{2504.03506}.

We consider the possibility that the modulus $\tau$ acts as the inflaton~\cite{2405.06497,2405.08924,2407.12081,2411.18603}, and/or evolves dynamically during inflation. 
In this case, the Standard Model Yukawa couplings acquire time-dependent phases throughout the inflationary epoch.
We show that this dynamics leads to enhanced cosmological collider signals (see~\cite{0911.3380, 1109.0292,1211.1624,1503.08043} for early works), as Standard Model particles effectively develop chemical potentials. 
As a consequence, the Higgs field acquires an inflationary vacuum expectation value.\footnote{The possibility that the Higgs field also participates in the inflationary dynamics in the context of modular inflation was discussed in~\cite{2504.01622}, but we do not consider this scenario here.}
Inflaton fluctuations then receive one-loop corrections from fermion loops
similar to the chiral anomaly.
We compute the resulting oscillatory contribution to the bispectrum.

\smallskip

In section~\ref{sec:model} we introduce the general framework. 
Given the length and technical nature of the computation, 
we begin in section~\ref{sec:outline} with an outline of the main steps and a preview of the result.
Details are provided in section~\ref{sec:detailed}, where we also formulate the precise quantization of 
fermions with general chemical potentials in de Sitter space, using the Dirac 4-component formalism. 
Our expressions extend related computations focused on axial chemical potentials~\cite{1805.02656,1908.00019,1907.10624,1910.12876,2010.04727}
while also correcting certain details.
As a result, our final result differs from previous computations.
Furthermore, as in~\cite{2304.13295,2605.21419} we go beyond estimates and
employ factorisation to compute the loop effect,
finding a result much smaller than claimed in earlier papers.
Fermion loops with chemical potentials have also been studied in the recent works~\cite{2611.23220,2611.20919,2611.31214}.
Conclusions are given in section~\ref{sec:concl}.

\section{The modular-invariant Standard Model}\label{sec:model}
We consider a minimal modular-invariant extension of the Standard Model (SM)~\cite{Ferrara:1989bc,Dixon:1990pc,Ibanez:1992hc,2501.16427,1706.08749}.
This adds a complex scalar, the modulus $\tau$, to the SM particle content, which comprises the $\SU(3)_c\otimes\SU(2)_L\otimes\U(1)_Y$ 
vectors $V=\{G,W,Y\}$, the Weyl fermions $\psi = \{L,E,Q,U,D,N\}$, and the Higgs doublet $\mathcal{H}$.
The extended Standard Model is invariant under the transformation
\beq \label{eq:modulartrans}
\tau\to \frac{a\tau+b}{c\tau+d},\qquad  \psi \to (c\tau+d)^{-k_\psi}\psi ,\qquad \mathcal{H} \to (c\tau+d)^{-k_H}\mathcal{H}
\eeq
with integers $a,b,c,d$ satisfying $ad - bc = 1$,
forming the $\SL(2,\mathbb{Z})$ modular group.
The coefficient $k_P$ is called the modular weight of the particle $P$.
Vectors are modular-invariant.
In this section $\tau$ is a dimensionless scalar.
The effective theory is described by the Lagrangian
\beq  \label{eq:Leff}
\Lag_{\rm eff} = \Lag_{\rm kin}+\Lag_{\rm Yuk}+\Lag_{\rm anom} - V(\mathcal{H},\tau) 
\eeq
where:
\begin{itemize}
\item $\Lag_{\rm Yuk} $ generalises the SM Yukawa terms, possibly including right-handed neutrinos $N$ and their masses $M$, 
by promoting the Yukawa couplings $Y$ to modular functions of $\tau$,
\beq
\Lag_{\rm Yuk} =-[  Y_{\rm u}(\tau) \,  QU\mathcal{H} + Y_{\rm d}(\tau) QD\mathcal{H}^*  + Y_{\rm e}(\tau) \,  LE\mathcal{H}^* + Y_\nu(\tau) LN\mathcal{H}+M(\tau) \frac{N^2}{2}+\hbox{h.c.} ] ,
\eeq
that ensure modular invariance by transforming as
$Y(\tau)\to (c\tau+d)^{k_Y} Y(\tau)$ with modular weights $k_{Y_{\rm u}} = k_Q + k_U  + k_H$, etc.
In theories with full modular invariance and no singularities
the Yukawas are given by Eisenstein functions, which make it possible to reproduce
the observed hierarchy of fermion masses and mixings (see e.g.~\cite{2305.08908}).

\item $\Lag_{\rm anom}$ contains anomalous terms whose coefficients are suppressed by one-loop factors,
\beq \label{eq:Lanom}
\Lag_{\rm anom}= 
\theta_3(\tau)   \frac{g_3^2}{32\pi^2} G^a_{\mu\nu} \tilde{G}^{a\mu\nu}+
\theta_2 (\tau) \frac{g_2^2}{32\pi^2} W^a_{\mu\nu} \tilde{W}^{a\mu\nu}+
\theta_1 (\tau)  \frac{g_Y^2}{16\pi^2} Y_{\mu\nu} \tilde{Y}^{\mu\nu} ,
\eeq
where $\theta_{1,2,3}$ must be appropriate modular functions of $\tau$ (see e.g. eq.~(15) of~\cite{2505.20395}).
In typical string compactifications, such anomalous terms are required to restore modular invariance. 
While the full string theory preserves modular invariance --- that arises as a subgroup of higher-dimensional reparametrization invariance --- this symmetry is generically anomalous in the low-energy QFT sector describing modes with sub-Planckian masses.

\item $\Lag_{\rm kin}$ contains the kinetic terms 
\beq\label{eq:Lkin}
\Lag_{\rm kin} = 
f^2 \frac{|\partial_\mu\tau|^2}{(-i\tau + i \bar\tau)^2} + 
 \sum_\psi \left[\frac{i}{2}\frac{\bar\psi \bar\sigma^\mu D_\mu\psi}{(-i\tau + i \bar\tau)^{k_\psi}} + \text{h.c.} \right]+
\frac{|D_\mu \mathcal{H}|^2}{(-i\tau + i \bar\tau)^{k_H}}  +\Lag_{\rm kin}^{\rm gauge} \eeq
where gauge vectors have standard kinetic terms.
In string models the modulus decay constant $f$ has a Planckian value $f = n \bp$ with integer $n$~\cite{Witten:1982hu}.
Here we consider sub-Planckian values, which can lead to detectable non-Gaussianities.
\end{itemize}
The kinetic terms are invariant under both gauge and modular transformations, thanks to the use of covariant derivatives 
$D_\mu$ that transform appropriately under both symmetries. 
In particular, modular invariance is ensured by introducing a minimal modular-covariant derivative, which takes the form
\beq \label{eq:Dcovmin}
D_\mu = \partial_\mu +i k \frac{ \partial_\mu \tau}{-i\tau+ i \bar\tau} \eeq
when acting on a field with modular weight $k$.

\smallskip

For our later purposes, the key non-trivial feature of the above theory
 is that the Yukawa couplings $Y(\tau)$ are complex functions of $\tau$.
 For example, the Eisenstein $E_4(\tau)$ is the unique  function with no poles
that transforms with weight 4.
The CP-invariant theory features a complex structure automatically provided by modular invariance.
Indeed, the modular transformation of matter fields in eq.\eq{modulartrans} induces a local U(1) phase rotation, which will be crucial in the following.
Mathematically, modular invariance can be interpreted as a sigma model on the coset $ \SL(2,\mathbb{R})/\U(1)\simeq  \SU(1,1)/\U(1) $, 
restricted to discrete transformations~\cite{Ferrara:1991uz}.
More general cosets with a U(1) stabiliser are expected to give rise to physics analogous to the modular effects discussed below.

\section{Outline of the effect}\label{sec:outline}
In this section we estimate the effect, outlining the essential physical mechanism.
From now on, we rescale the modulus field to have canonical mass dimension 1
and a canonical kinetic term around its inflationary vacuum expectation value.
We focus on the real part of $\tau$, which controls CP violation.
The canonical $\Re\tau$ appears in the modular covariant derivative with decay constant $f$ for any $\Im\tau$.

\subsection{Choosing a basis in field space}
Two alternative field redefinitions allow us to partially remove the 
modulus $\tau$ from the Lagrangian of eq.\eq{Leff}, thereby simplifying the description of the system: 
\begin{enumerate}
\item[$K$)] An appropriate rephasing of the Higgs $\mathcal{H}$ and fermion $\psi$ fields
allows us to choose a basis in field space that eliminates $\tau$ from the matter kinetic terms,
thereby removing all couplings of $\partial_\mu \tau$  to particle currents $J^\mu_P$.
In this basis the matter fields carry vanishing effective modular charge, and
$\tau$ remains as a physical scalar that cannot be eliminated
in the Yukawa couplings and in the loop-suppressed anomalous terms.

\item[$Y$)] An alternative rephasing achieves the opposite: it removes $\tau$ from the Yukawa terms 
(and partially from the anomalous terms).
In this basis the canonical $\partial_\mu\tau$ couples to the various particle currents $J^\mu_P$
with strength $k'_P/f$, parameterised by an effective weight $k'_P$.
These are given by the original modular weights $k_P$, shifted by combinations of the derivatives
$d\arg Y(\tau)/d\tau$ of the phases of the various Yukawa couplings $Y(\tau)$,
and can be computed in any specific model.
\end{enumerate}
During inflation, $\tau_0(t)$ is time-dependent; the subscript ``0'' denotes the background.
As a result, massive particles $P$ acquire either a mass with time-dependent phase in the $K$ basis,
or a chemical  potential $\mu_P \sim k'_P \dot\tau_0/f$ in the ~$Y$ basis.
A combination of both arises in a generic basis.
Physical observables are basis-independent.
We will compute in the $Y$ basis, where the free particle dynamics is simpler and the remaining interactions can be treated perturbatively. 

\smallskip

The physics described above resembles that of models introduced ad hoc 
to enhance cosmological collider signals through chemical potentials, featuring a complex scalar
and a U(1)-breaking term that renders the associated phase physical~\cite{2010.04727,2507.22978}.
A breaking term {\em linear} in scalar fields gives tree-level effects~\cite{2010.04727}.
A breaking term {\em quadratic} in the fields generates one-loop effects~\cite{2507.22978}.
In our modular theory, the U(1)-breaking Yukawa interactions are {\em cubic} in the matter fields.
So, each Yukawa induces a two-loop contribution to non-Gaussianities (in the $\SU(2)_L$-preserving phase), mediated by the SM fermions and by the Higgs.
Although enhanced by chemical potentials, such two-loop effects remain small.
A relatively larger one-loop contribution arises from right-handed neutrinos, as their mass term $M(\tau) N^2/2$ 
is quadratic in the matter fields.
A similar one-loop effect also arises from Standard Model fermions, because 
$\SU(2)_L$ gets broken during inflation, as we now discuss.

\subsection{Higgs mass during inflation}
The Higgs is a special scalar that happens to have a small weak-scale mass $M_H \ll \bp$  at the minimum of the SM potential.
As a result, corrections to its squared mass during inflation can be particularly significant. 
Several contributions are generically present:
\begin{enumerate}
\item Expanding the modular-covariant Higgs kinetic term in eq.\eq{Lkin} at second order in $\tau$ reveals a negative contribution 
to the squared Higgs mass, as typical in the presence of a chemical potential $\mu_H$:
\beq \delta M_H^2 = -\mu_H^2,\qquad
\mu_H \sim k'_H \frac{\dot\tau_0}{f}
\label{eq:deltaMh}.\eeq

\item A non-minimal coupling of the Higgs to the curvature $R$, described by a  $-\xi_H |\mathcal{H}|^2 R$ interaction in the Lagrangian,
induces a Hubble-scale Higgs mass, 
\beq \delta M_H^2 = (\xi_H + 1/6) R= -12 (\xi_H + 1/6) H^2.\eeq
\item The Higgs mass squared receives
a positive correction $\delta M_H^2 \sim g^2 T^2$ in a thermal bath.
We assume a negligible temperature $T \ll H$ during inflation.
\item A model-dependent direct coupling of the inflaton $\tau$ to the Higgs can also 
generate a significant contribution to the squared Higgs mass.
In modular theories both $M_H^2$ and $\xi_H$ are given by $(\Im \tau)^{-k_H}$ times a modular invariant function, such
as a constant. In such a case the direct-coupling effect is negligible.


\end{enumerate}
We assume that the effect of eq.\eq{deltaMh} dominates. 
Taking into account the terms linear and quadratic in $\mu_H$, the Higgs dispersion relation 
is $(E+\mu_H)^2=k^2 + M_H^2$.
So, during inflation,
the Higgs forms a Bose--Einstein condensate, acquiring a large vacuum expectation value $v= \med{|\mathcal{H}|^2}^{1/2}= \mu_H/\sqrt{2\lambda_H}$, kept finite by the Higgs quartic $\lambda_H$.
In the SM, the quartic Higgs coupling runs to small values $\lambda_H \sim 0.01$ at high energy, and possibly turns negative (see e.g.~\cite{1307.3536}).
We ignore the possibility of a negative Higgs quartic.
Furthermore we  assume that Higgs fluctuations are heavier than 
$3H/2$, so that the Higgs does not accumulate inflationary quantum fluctuations.
Different directions of $\med{\mathcal{H}}$ in different Hubble patches correspond to small electromagnetic fields~\cite{Vachaspati:1991nm}.

\medskip

As a result of the large inflationary Higgs vacuum expectation value $v \sim \mu_H$,
the SM fermions acquire inflationary Dirac mass terms 
$ m \bar\Psi\Psi + \hbox{h.c.}$ where $\Psi = (\psi_L, \bar\psi_R)$ and $m=y v$.
The Yukawa couplings $y$ of SM fermions range from $10^{-6}$ to 1.
Furthermore, the time dependence of the modulus contributes to
chemical potentials 
\beq \label{eq:muLR}\mu_{{L,R}} = k'_{\psi_{L,R}} \dot\tau_0/f\eeq
for the left-handed and right-handed components of each SM fermion.
While the chemical potential of a scalar just shifts its squared mass~\cite{2010.04727,2507.22978},
chemical potentials have more significant effects on fields with spins, in our case fermions.
In flat space, the fermion dispersion relation becomes
\beq \label{eq:Ek} (E_{kh} +\mu_V)^2 = (h k - \mu_A)^2 + m^2\eeq
where $\mu_V = (\mu_R+\mu_L)/2$ and $\mu_A =(\mu_R-\mu_L)/2$
are the vector and axial chemical potentials, and
 $h=\pm 1$ is the helicity.
In a de Sitter inflationary background the momentum redshifts as 
$k = k_{\rm comoving}/a$,
leading to enhanced fermion production when $k_{\rm comoving}/a\sim \mu_A$~\cite{1803.04501,1910.12876}.
Indeed, in a Bogolyubov computation, a particle mode with momentum $k$ starts at early times with $\beta_k\simeq 0$.
It gets populated at this point,
as the adiabatic suppression factor $e^{-E^2_{kh}/\dot E_{kh}}$ 
becomes of order unity 
in a $\Delta k$ range of order $m$.
The final abundance is $|\beta_k^2| \simeq e^{-\pi m^2/H\mu_A}$,
which equals the Fermi--Dirac value of unity for masses below the threshold for exponential Boltzmann suppression.
A positive $\mu_A>0$ produces fermions with positive helicity $h=1$,
while a negative $\mu_A$ produces $h=-1$.
The total fermion number density $n \sim 4\pi m \mu_A^2 e^{-\pi m^2/ H \mu_A}$.
This enhanced fermion number density is claimed to induce a $\mu_A^2$
enhancement in non-Gaussianities of cosmo-collider type,
as it controls the long-range propagation 
of the two soft fermion lines that generate the
cosmological collider signal (see e.g.~\cite{1508.00891,1803.04501}).\footnote{Similar physics was considered in~\cite{1908.00019,1907.10624}, assuming a large chemical potential $\mu_t$ for top quarks, 
which induces a one-loop contribution $\delta M_H^2 \sim -y_t^2 \mu_t^2/(4\pi)^2$ to the Higgs mass squared.
We ignore this loop effect, as the modular theory directly provides a larger tree-level squared Higgs mass, eq.\eq{deltaMh}.}


\subsection{Non-Gaussianities}
During inflation driven by a homogeneous inflaton background $\tau_0(t)$,
quantum fluctuations $\delta\tau$ source curvature perturbations $\zeta = -H \delta \tau /\dot\tau_0$.
This leads to the scalar power spectrum 
$P_\zeta = (H^2/2\pi\dot\tau_0)^2 = (H^2/8\pi^2 \bp^2 \epsilon)$
where $\epsilon$ is the slow-roll parameter.
On currently observable cosmological scales, the power spectrum is measured to be nearly scale-invariant, with $P_\zeta \approx 2~10^{-9}$, which
is reproduced for $\dot\tau_0\approx (60 H)^2$.
The maximal inflationary Hubble scale allowed by the bound
$r = 16 \epsilon \lesssim 0.032$ on the tensor-to-scalar ratio 
is $H = \pi \bar{M}_{\rm Pl} \sqrt{r P_\zeta/2} \lesssim 2~10^{-5}\bar{M}_{\rm Pl}$.
This means that chemical potentials $\mu\sim\dot\tau_0/f$ are larger than $H$ 
if the modulus decay constant $f$ has a sub-Planckian value,
\beq f \lesssim 60^2 H \lesssim 0.07\bp.\eeq
On the other hand, unitarity demands $f \gtrsim \dot\tau_0^{1/2}$~\cite{1805.02656}
restricting chemical potentials to be $\mu \lesssim \dot\tau_0^{1/2}\sim 60 H$.

The bispectrum of curvature perturbations $\zeta_k$ is usually parameterised in terms of a dimensionless shape function $S$ as
\beq \med{\zeta_{\vec k_1}\zeta_{\vec k_2}\zeta_{\vec k_3}} =(2\pi)^3 \delta(\vec{k}_1+\vec{k}_2+\vec{k}_3)
 \frac{(2\pi)^4 P_\zeta^2}{(k_1 k_2 k_3)^2} S \bigg( \frac{k_1}{k_3}, \frac{k_2}{k_3}\bigg) 
 \simeq - \left(\frac{H}{\dot{\tau}_0}\right)^3\med{\delta\tau_{k_1}\delta\tau_{k_2}\delta\tau_{k_3}}.
\eeq
The oscillatory cosmological collider signal appears
in the squeezed limit $ k_3\ll k_1\approx k_2   $, where
the momenta are usually referred to as short and long
\beq k_L \equiv k_3 \ll k_{1} \approx k_2 \equiv k_S.\eeq
In this limit 
the shape function $S$ contains a smooth term and an  oscillatory term in $\ln k_L/k_S$.
The effect arises from fermion production and absorption~\cite{2112.03448}.
Fermion-antifermion pairs are initially produced gravitationally with number density $n $ and energy $\mu$.
The fermions later annihilate, producing the clock signal
$(k_L/k_S)^{2 i \lambda}$ times the $(k_L/k_S)^2$ scaling from
the inflationary dilution of their densities (see e.g.~\cite{1612.08122}).
We define the dimensionless combinations 
\beq \tilde{m} \equiv \frac{m}{H},\qquad \tilde\mu_{A,V} \equiv \frac{\mu_{A,V}}{H},\qquad
\lambda \equiv   \sqrt{\tilde{m}^2 + \tilde\mu_A^2}=\frac{\sqrt{m^2 + \mu_A^2}}{H} 
\eeq
The contribution to a fermion loop in our theory can be written as
\beq \label{eq:shapefNL}
S \simeq f_{\rm NL}^{\rm osc} \bigg(\frac{k_L}{k_S}\bigg)^{2+2i \lambda}+\hbox{h.c.} +\hbox{(non-oscillatory~terms}).\eeq
Our goal is to compute the $f_{\rm NL}^{\rm osc}$ parameter from a fermion loop.
Previous estimates and computations in models with axial chemical potentials found~\cite{1805.02656,1908.00019,1910.12876}
\beq \label{eq:fNLapprox}
f_{\rm NL}^{\rm osc} \approx 
\frac{1}{2\pi \sqrt{P_\zeta}} \frac{1}{(4\pi)^2} \bigg(\frac{m}{f}\bigg)^3  \frac{H}{m}
\frac{n}{H^3}  \approx \pi
P_\zeta \tilde{\mu}^5_A  \tilde{m}^3 e^{-\pi \tilde{m}^2/\tilde\mu_A} .
\eeq
The first term is the conventional normalization of $f_{\rm NL}$.
The second term is a typical one-loop factor.
The $m/f$ vertex factor for each interaction arises because a massless fermion would be decoupled due to conformal invariance.
The second expression in eq.\eq{fNLapprox} is obtained by substituting $1/f =\mu/\dot\tau_0 =2\pi\mu_A \sqrt{P_\zeta}/H^2$,
and contains a significant $\tilde\mu^5_A$ enhancement from the axial chemical potential.
$f_{\rm NL}^{\rm osc}$ remains small if the modulus decay constant is $f \approx \bp$, as chemical potentials are small.
A large $f_{\rm NL}^{\rm osc}$ arises if the inflationary fermion mass is a few times the Hubble scale,
\beq \tilde m \approx \frac{\mu_H}{H}\frac{ y}{\sqrt{2\lambda_H}} \sim \hbox{few}\eeq
and if $\mu_A$ is tens of times higher.
A large $\tilde\mu_A$ requires $f \approx 60 H$, around the unitarity limit~\cite{1805.02656}.
In the SM at energies around $10^{13}\GeV$ one finds $y_b/\sqrt{2\lambda_H}\approx 0.05$ and $y_t/\sqrt{2\lambda_H}\approx 3$,
as $\lambda_H$ is accidentally small and significantly uncertain  (see e.g.~\cite{1307.3536}).
So the condition on the mass can be satisfied for the bottom quark, for order unity values of modular weights,
and for the top quark, if fermion modular weights are larger than the Higgs modular weight.
No collider constraints apply, as plausible values of the
inflationary scale $H$ are expected to be much higher than the energy scales probed by current colliders.


However, our final result in eq.\eq{fNLasy} will be smaller than eq.\eq{fNLapprox}.

%

\begin{figure}[t]
$$\includegraphics[width=\textwidth]{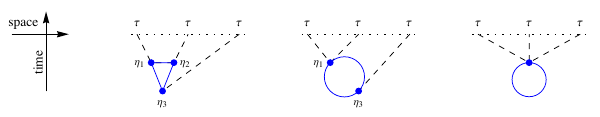}$$
\begin{center}
\caption{\em\label{fig:ModularCosmoFeyn}
Cosmo-collider diagrams for one-loop corrections to the inflaton modulus bispectrum.
In the first diagram two fermions are produced at an early time $\eta_3$ and later annihilate at $\eta_{1,2}$.
}
\end{center}
\end{figure}

\section{Detailed computation}\label{sec:detailed}

Cosmological collider signals can be systematically computed using the
Schwinger-Keldysh formalism~\cite{1703.10166}. The enhancement discussed above gets encoded in 
fermion propagators.


\subsection{A fermion with chemical potentials in de Sitter}
We consider a 4-component Dirac fermion $\tilde\Psi$ with generic vector and axial chemical potentials in de Sitter,
with metric $ds^2 = a^2 (\eta)(d\eta^2 - d\vec{x}\,^2)$, 
where $a=-1/H\eta$ is the scale factor
and $\eta$ is conformal time. The two-component formulation for a Weyl fermion with an axial chemical potential in de Sitter space can be found in~\cite{1805.02656,1907.10624, 2304.09428}.
The action for the Dirac fermion is given by
\beq S =  \int d^3x\, d\eta \, \bar{\tilde\Psi} \bigg[a^3 ( i\slashed{D}  + \slashed{V}+ \slashed{A}\gamma^5) - a^4 m \bigg]\tilde\Psi .\eeq
The chemical potentials are written, in covariant notation,
as the time components of a vector, $V_\eta = a \mu_V$,
and an axial vector, $A_\eta=a \mu_A$.
The gravitational covariant derivative $D_\mu$ reduces to its flat space form
after performing the Weyl rescaling $\tilde\Psi = a^{-3/2}\Psi$, 
\beq S =  \int d^3x\, d\eta\, \bar{\Psi} \bigg[i\slashed{\partial}  + 
\slashed{V}+ \slashed{A}\gamma^5 - a m \bigg]\Psi .\eeq
We use the chiral basis
\beq
\gamma ^\mu= \begin{pmatrix}  0&\sigma^\mu \\  \bar\sigma^\mu&0 \end{pmatrix} ,\qquad
\gamma ^5 =  i \gamma^0\gamma^1\gamma^2\gamma^3=  \begin{pmatrix} -1&0\cr 0& +1 \end{pmatrix} \eeq
where $\sigma^\mu = (1,\sigma^i)$, $\bar\sigma^\mu = (1,-\sigma^i)$,
and 1 denotes the $2\times 2$ unit matrix.
The Dirac fermion can be written in terms of two Weyl fermions as $\Psi=(\psi_L,\bar \psi_R)$.
It can optionally be reduced to a single 2-component 
Weyl spinor $\psi_L = \psi_R$ by imposing the Majorana reality condition,
in which case the vector chemical potential vanishes.

\subsubsection{Fermion modes in de Sitter}
The Dirac equation in de Sitter is
$[ i \slashed{\partial} + \slashed{V} +  \slashed{A}\gamma^5 - am ]\Psi=0$.
After Fourier transforming in space, the mode functions of $\eta,\vec{k}$ satisfy
\beq [\gamma^0 (i \partial_\eta+a\mu_V + a \mu_A \gamma^5) - \vec\gamma\cdot\vec k - a m]\Psi = 0.\eeq
To solve this matrix equation, we diagonalise $\gamma^5$, obtaining
\beq \label{eq:DiraceqLR}
\begin{pmatrix}  -am & i \partial_\eta + a \mu_R - \vec{k}\cdot\vec\sigma \\ 
 i \partial_\eta + a\mu_L + \vec{k}\cdot\vec{ \sigma}&-am \end{pmatrix}
\begin{pmatrix}  \psi_L \\ \bar \psi_R\end{pmatrix}=0
\eeq
where $\mu_L = \mu_V-\mu_A$ and $\mu_R = \mu_V+\mu_A$.
So
\beq\begin{aligned}
[i\partial_\eta+a \mu_L  + \vec\sigma\cdot\vec k ] \psi_L = am \bar{\psi} _R,\cr
[i\partial_\eta+a \mu_R  -  \vec\sigma\cdot\vec k ] \bar\psi_R = am {\psi} _L .
\end{aligned}\eeq
To solve these equations, it is convenient to further diagonalise $(\vec\sigma\cdot\vec k) \chi_h = hk \chi_h$ in the helicity basis $h=\pm 1$.
Expanding $\psi_L = \sum_h \psi_{Lh}  \chi_h$ and $\bar\psi_R = \sum_h \bar\psi_{Rh}  \chi_h$, 
the Dirac equation becomes
\beq\begin{aligned}\label{eq:Diraccoupled}
[i\partial_\eta+a \mu_L  + h k ] \psi_{Lh} = am \bar{\psi} _{Rh},\cr
[i\partial_\eta+a \mu_R  -  hk] \bar\psi_{Rh} = am {\psi} _{Lh} .
\end{aligned}\eeq
Disentangling the two equations gives
\beq\begin{aligned}
\psi''_{Lh} - aH (1+2i \tilde\mu_V)\psi'_{Lh}+\bigg[ (hk-a H \tilde\mu_A)^2 + a^2 (m^2 -H^2 \tilde\mu_V^2) + i a h k H  \bigg]\psi_{Lh} = 0, \cr
\bar\psi''_{Rh} - aH (1+2i \tilde\mu_V)\bar\psi'_{Rh}+\bigg[ (hk-a H \tilde\mu_A)^2 + a^2 (m^2 -H^2 \tilde\mu_V^2) - i a h k H  \bigg]\bar\psi_{Rh} = 0
\end{aligned}\eeq
where $\tilde\mu_V  =\mu_V/H$ and $\tilde\mu_A = \mu_A/H$
are dimensionless chemical potentials and the prime here denotes the derivative with respect to conformal time $\eta$.
These equations are solved by Whittaker functions $W$ as
\beq \begin{aligned} 
\psi_{Lh} &  \propto (-2 k\eta)^{-1/2 - i \tilde\mu_V} 
[c_+ W_{-h(1/2+ i \tilde\mu_A), i \lambda}(2i  k\eta) + c_- W_{+h(1/2+ i \tilde\mu_A), i \lambda}(-2i  k\eta)]
,\cr
\bar\psi_{Rh}& \propto  (-2 k\eta)^{-1/2 - i \tilde\mu_V} [c_+ W_{+h(1/2- i \tilde\mu_A), i \lambda}(2i  k\eta)+
c_-  W_{-h(1/2- i \tilde\mu_A), i \lambda}(-2i  k\eta)]
\end{aligned}\eeq
where $c_+$ ($c_-$) multiplies a mode with initial positive (negative) frequency,
$h^2=1$, $\lambda=\sqrt{m^2 + \mu_A^2}/H$, $\tilde m = m/H$,
and the vector chemical potential just multiplies the solution  by a  $\eta^{- i \tilde\mu_V}$ factor.

The Bunch--Davies particle solutions that satisfy the coupled eq.\eq{Diraccoupled} and are normalised such that
$|u_{Lh}|^2 + |\bar u_{R h}|^2=1$ are
\beq \begin{aligned}\label{eq:psiLR+-}
u_{\vec  k,+} = 
\begin{pmatrix} u_{L+}  \chi_+ \cr\bar u_{R+}  \chi_+ \end{pmatrix}  &=
(-2 k\eta)^{-1/2 - i \tilde{\mu}_V} e^{+\pi  \tilde{\mu}_A/2}
\begin{pmatrix}   \tilde{m}    W_{-1/2-i \tilde{\mu}_A, i \lambda}(2i  k\eta)   \chi_+ \cr
   i   \, W_{+1/2- i \tilde{\mu}_A, i \lambda}(2i  k\eta) \chi_+ 
\end{pmatrix} , 
\cr
u_{\vec  k,-} = \begin{pmatrix} u_{L-} \chi_- \cr\bar u_{R-} \chi_- \end{pmatrix}  & =
(-2 k\eta)^{-1/2 - i \tilde{\mu}_V} e^{-\pi  \tilde{\mu}_A/2} 
\begin{pmatrix} i W_{+1/2+ i \tilde{\mu}_A, i \lambda}(2i  k\eta) \chi_-  \cr 
\tilde m  W_{-1/2+ i \tilde{\mu}_A, i \lambda}(2i  k\eta) \chi_- \end{pmatrix}.
\end{aligned}\eeq
The corresponding antiparticle solutions are
\beq \begin{aligned}\label{eq:vLR+-}
v_{\vec  k,+} = 
\begin{pmatrix} v_{L+} \chi_+ \cr\bar v_{R+} \chi_+ \end{pmatrix}  &=
(-2 k\eta)^{-1/2 - i \tilde{\mu}_V} e^{+\pi  \tilde{\mu}_A/2}
\begin{pmatrix}   i   W_{+1/2+ i \tilde{\mu}_A, i \lambda}(-2i  k\eta)  \chi_+ \cr
   \tilde{m}   \, W_{-1/2+ i \tilde{\mu}_A, i \lambda}(-2i  k\eta) \chi_+ 
\end{pmatrix}  , \cr
v_{\vec  k, -} = \begin{pmatrix} v_{L-}  \chi_- \cr\bar v_{R-} \chi_-\end{pmatrix}  & =
(-2 k\eta)^{-1/2 - i \tilde{\mu}_V} e^{-\pi  \tilde{\mu}_A/2} 
\begin{pmatrix} \tilde{m}  W_{-1/2- i \tilde{\mu}_A, i \lambda}(-2i  k\eta) \chi_-  \cr 
i  W_{+1/2-  i \tilde{\mu}_A, i \lambda}(-2i  k\eta)\chi_- \end{pmatrix}.
\end{aligned}\eeq
These spinors satisfy the completeness relation
\beq \label{eq:completeness}
\sum_h (u_h \otimes u_h^\dagger + v_h\otimes v_h^\dagger) = 1\eeq
in view of
\beq \chi_h \otimes \chi_h^\dagger = \Pi_h = \frac{1+ h \vec\sigma\cdot\hat{k}}{2},
\qquad
\sum_h   \chi_h \otimes \chi_h^\dagger = 1 
\eeq
where $\hat{k}\equiv \vec{k}/k$, and of the Whittaker identities, 
$W_{1/2-i \tilde{\mu}_A, i \lambda}(2i k\eta)^* = W_{1/2+i \tilde{\mu}_A, i \lambda}(-2i k\eta)$,
\beq (-2k\eta)^{-1} e^{\pm\pi\tilde{\mu}_A} [|W_{1/2+i\tilde{\mu}_A, i \lambda}(\mp 2 i k\eta) |^2 + \tilde{m}^2 |W_{-1/2-i\tilde{\mu}_A, i \lambda}(\pm 2 i k\eta) |^2 ]=1.
\eeq
The Whittaker functions describe Bogolyubov creation of quanta from the de Sitter expansion.

In the limit of vanishing chemical potentials $\mu_{A,V}\to 0$   one 
has $u_{L-}=\bar u_{R+}$, $\bar u_{R-}= u_{L+}$, etc.
So one can focus on the states with helicity $h=+1$, 
and rewrite their mode functions in terms of Hankel functions,
finding agreement with~\cite{2309.10841}\footnote{We thank Z.\ Qin and Y.\ Zhu for pointing out this formula.}
\beq \begin{aligned} 
W_{-\frac{1}{2},\mu}(-2iz)
=
\frac{\sqrt{\pi }z}{2{\mu}}
e^{(2\mu+1)\pi i/4}
\left[
H^{(1)}_{\mu+\frac{1}{2}}(z)
+
i\, H^{(1)}_{\mu-\frac{1}{2}}(z)
\right],
\cr
W_{\frac{1}{2},\mu}(-2iz)
=
\frac{\sqrt{\pi} z}{2}
e^{(2\mu-1)\pi i/4}
\left[
H^{(1)}_{\mu-\frac{1}{2}}(z)
+
i\, H^{(1)}_{\mu+\frac{1}{2}}(z)
\right].
\end{aligned}\eeq
In the limit $m,\lambda\to 0$ one recovers the flat-space  result, e.g.\ $u_{\vec{k},+} = (-1)^{1/4} (0,e^{-i k\eta} \chi_+)$
and the dispersion relation of eq.\eq{Ek}.

\subsubsection{Fermion field}
The (rescaled) fermion field, quantised in the Bunch--Davies vacuum and satisfying
$\{\Psi(\eta,\vec{x}),\Psi^\dagger(\eta,\vec{x}')\} = \delta(\vec{x}-\vec{x}') $,
can be written as
\beq\Psi (\eta,\vec{x})= \int\frac{d^3k}{(2\pi)^3}e^{+i \vec{k}\cdot\vec{x}}
\sum_{h=\pm1} [u_{\vec k,h} b_{\vec k,h}+ v_{\vec  k,h} d_{\vec  k,h} ^\dagger]
=\int\frac{d^3k}{(2\pi)^3}e^{+i \vec{k}\cdot\vec{x}}\Psi(\eta,\vec{k}).
\eeq
The creation and annihilation operators anticommute as
\beq\label{eq:anticom}
\{b_{\vec k,h}, b_{\vec k',h'}^\dagger\} =
\{d_{\vec k,h}, d_{\vec k',h'}^\dagger\} =
(2\pi)^3 \delta(\vec{k}-\vec{k}')\delta_{hh'}.\eeq
The conjugate field is
\beq\bar{\Psi} (\eta,\vec{x})= \int\frac{d^3k}{(2\pi)^3}e^{-i \vec{k}\cdot\vec{x}}
\sum_{h=\pm1} [\bar{u}_{\vec  k,h} b^\dagger_{\vec k,h}+ \bar{v}_{\vec  k,h} d_{\vec  k,h} ]
=\int\frac{d^3k}{(2\pi)^3}e^{-i \vec{k}\cdot\vec{x}} {\bar\Psi} (\eta,\vec{k}).\eeq
Imposing the Majorana condition $\Psi^c\equiv C \bar\Psi^T = \Psi $ would force $\mu_V=0$ and identify $b_{\vec{k},h}=d_{-\vec{k},h}$,
$v_{\vec{k},h}=C \bar{u}_{-\vec{k},h}^T$
with helicity-spinor phases chosen so that $\epsilon\chi_\pm^*(-\vec{k})=\mp \chi_\pm(\vec k)$.


\subsection{Schwinger-Keldysh fermion propagators}
The SK formalism performs a closed-time-contour path integral by doubling the fields into $\Psi_a$
with $a=\pm1$, such that the generating functional is
\begin{align}
Z[\eta_a,\bar\eta_a]
&=\int \prod_{a=\pm}\mathcal D\Psi_a\,\mathcal D\bar\Psi_a\;
\exp\Bigg\{
i\sum_{a=\pm} a\,  \bigg[ S[\Psi_a,\bar\Psi_a]
+  \int d^4x\,
\big[\bar\eta_a(x)\Psi_a(x)+\bar\Psi_a(x)\eta_a(x)\big] \bigg]
\Bigg\}
\end{align}
with fields glued as $\Psi_+(t_f, \vec{x}) = \Psi_-(t_f,\vec{x})$.
In this way vertices give $i \sum_a a$ factors, and the SK propagators are
\begin{align}
D_{ab}(x,y)
&=
\left.
\frac{\delta}{i a\,\delta \bar\eta_a(x)}
\frac{\delta}{i b\,\delta \eta_b(y)}
Z[\eta]
\right|_{\eta=0} = -i\,\langle {\rm T}_{\rm cl} \,\Psi^a(x)\,\bar\Psi^b(y)\rangle .
\end{align}
Here the $-i$ arises from Gaussian integration,
and ${\rm T}_{\rm cl} $ is the closed-contour time ordering.
So
\beqa \label{eq:DSK}
D_{-+} (\vec{k},\eta_1, \eta_2)&= -i \med{\Psi(\eta_1, \vec{k}) \bar{\Psi}(\eta_2,\vec{k})}=
- i \sum_h u_{\vec k,h}(\eta_1) \otimes \bar u_{\vec{k},h}(\eta_2) ,
\cr
D_{+-} (\vec{k},\eta_1, \eta_2)& =
 -i \med{ {\rm T}_{\rm cl} \Psi_+ (\eta_1, \vec{k}) \bar{\Psi}_- (\eta_2,\vec{k})} = 
 i \med{\bar{\Psi}(\eta_2, \vec{k}) \Psi(\eta_1,\vec{k})}=
 i \sum_h   v_{\vec{k},h}(\eta_1) \otimes \bar v_{\vec k,h}(\eta_2)   ,\cr
D_{++}(\vec{k},\eta_1, \eta_2)&=
 -i \med{{\rm T}\,\Psi(\eta_1, \vec{k}) \bar{\Psi}(\eta_2,\vec{k})}= \theta(\eta_1-\eta_2) D_{-+} (\vec{k},\eta_1, \eta_2)  +
\theta(\eta_2 - \eta_1) D_{+-}(\vec{k},\eta_1, \eta_2),\cr
D_{--}(\vec{k},\eta_1, \eta_2)&=- i \med{\bar{\rm T}\, {\Psi}(\eta_1, \vec{k}) \bar{\Psi}(\eta_2,\vec{k})} 
= \theta(\eta_1-\eta_2) D_{+-}  (\vec{k},\eta_1, \eta_2) + \theta(\eta_2 - \eta_1) D_{-+}(\vec{k},\eta_1, \eta_2).
\eeqa

These propagators are used below keeping complete fermion modes.
Their late-time limit, used in previous computations, is discussed in appendix~\ref{app:late}.

\subsection{Interaction vertices}
We expand the inflaton modulus $\tau$ in small fluctuations as
$\tau = \tau_0(\eta) + \delta \tau (\eta,\vec{x})$,
where $\eta$ is conformal time.
The interaction of each SM particle $P$ with the modulus
can be expanded in powers of the small fluctuation $\delta\tau$ as
\beq \label{eq:interactions}
f_P(\tau)(\partial_{\mu} \tau) J^\mu_P \simeq 
f _{P0} \dot\tau_0 J^0_P +
[f_{P0} \,\partial_\mu \delta\tau\, J^\mu_P + f' _{P0} \dot\tau_0 \delta\tau J^0_P]+
\bigg[f'_{P0} \delta\tau \,\partial_\mu \delta\tau\, J^\mu_P + f''_{P0} \dot\tau_0 \frac{\delta\tau^2}{2}J^0_P\bigg]
+
{\cal O}(\delta\tau)^3.\eeq
The first term with $f_{P0} = f_P(\tau_0)$ provides a chemical potential $\mu_P = f_{P0} \dot\tau_0$
to each particle $P$.
Describing SM fermions as Dirac fermions, the chemical potentials $\mu_L$ and $\mu_R$
of their left-handed and right-handed components combine into vector and axial chemical potentials
as described below eq.\eq{DiraceqLR}, where
the fermion equations of motion were solved by approximating the chemical potentials as constant
and the inflationary spacetime as de Sitter.

\medskip

\begin{figure}[t]
$$\includegraphics[width=\textwidth]{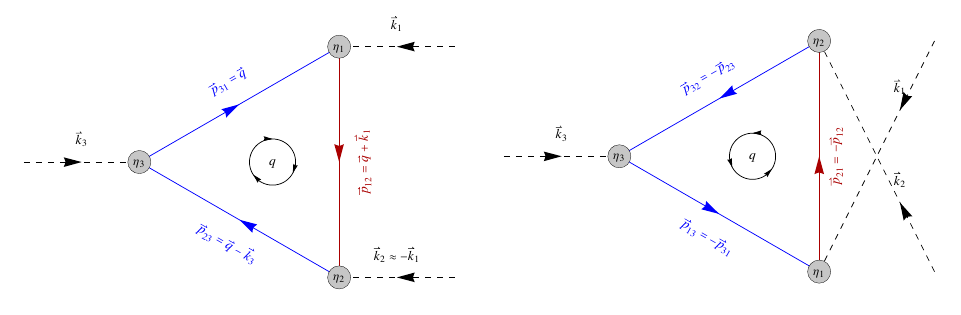}$$
\begin{center}
\caption{\em Dominant one-loop Feynman diagram.
We show the momenta flowing in the triangle fermion loop.
The hard 12 propagator is denoted in red; the soft 23 and 13 propagators are denoted in blue.
\label{fig:Feyn2}}
\end{center}
\end{figure}

The first vertex, linear in $\delta\tau$ and involving a Dirac fermion, is dominant and can be written as
\beq \Gamma^\mu = \gamma^\mu \frac{\mu_V+ \mu_A \gamma^5}{\dot\tau_0}\eeq
where we have used the canonically normalised fermion $\Psi$.
This vertex induces a loop correction from the anomaly-like triangle fermionic diagram in  fig.\fig{Feyn2}.
All other vertices  in eq.\eq{interactions} can be neglected as long as $f_P$ remains nearly constant during one $e$-fold.
This happens in theories where the Yukawa phases smoothly undergo an order-unity variation during the $N\gtrsim 60$ $e$-folds of inflation,
while varying by only $\sim 1/N$ during one $e$-fold.
For example, the second vertex linear in $\delta\tau$ gives a contribution suppressed by $\dot{f}_{P0}/H f_{P0} $,
in view of $\partial_\mu \delta \tau \sim H \delta \tau$ around horizon exit.
The first quadratic vertex leads to a similar suppression, 
and the second quadratic vertex leads to a $\ddot f_P / H^2 f_P$ suppression.
These $1/N$ suppressions can be seen as slow-roll suppressions in inflationary models where the slow-roll parameters are $\epsilon\sim \eta\sim 1/N$.

\subsection{Computation of the loop correction}
The triangle diagram in fig.\fig{Feyn2} contributes to the three-point function of $\delta\tau$ as
\beq\label{eq:tau3}
\med{\delta\tau_{\vec k_1}\delta\tau_{\vec k_2}\delta\tau_{\vec k_3}}'
= -\int_{-\infty}^0 d\eta_{1,2,3}
\!\!\! \sum_{a_1,a_2,a_3=\pm1} \!\!\!\!
a_1a_2a_3 \,
F_{\mu_1 a_1} (\vec k_1,\eta_1)
F_{\mu_2 a_2}(\vec k_2,\eta_2)
F_{\mu_3 a_3 } (\vec k_3,\eta_3)\int \frac{d^3q}{(2\pi)^3} A^{\mu_1\mu_2\mu_3}_{a_1 a_2 a_3}.
\eeq
Here the $'$ indicates that we have factored out the momentum-conserving factor $(2\pi)^3 \delta(\vec{k}_1+\vec{k}_2+\vec{k}_3)$.
The SK indices are $a_{1,2,3}$.
The external scalar wave functions are~\cite{1805.02656} 
\begin{align}
F_{\mu {a}}(\vec{k}, \eta) \equiv 
e^{i \vec{k}\cdot \vec{x}} \partial_\mu (G_a
e^{-i \vec{k}\cdot \vec{x}})=
\binom{\partial_{\eta} G_{a}}{-i\vec{k} G_{a}}=\frac{H^2}{2 k^3}\binom{k^2 \eta}{-i\vec{k}(1-i{a} k \eta)} e^{i{a} k \eta}
\end{align}
where $G_{ {a}}= e^{  {i}{a} k \eta} H^2 (1 - {i}{a} k \eta) /{2 k^3}$.
The term inside the $d^3q$ loop integral is the fermion trace 
\beq A^{\mu_1\mu_2\mu_3}_{a_1 a_2 a_3} =  N_c \Tr [
\Gamma^{\mu_1} D_{a_1a_2}(\vec p_{12},\eta_1, \eta_2)
\Gamma^{\mu_2} D_{a_2a_3}(\vec p_{23},\eta_2,\eta_3)
\Gamma^{\mu_3} D_{a_3a_1}(\vec p_{31},\eta_3,\eta_1) ]+ (1\leftrightarrow 2)\eeq
with $N_c=1$ for a lepton and $N_c=3$ for a quark.
Here $1\leftrightarrow 2$ means exchanging particles 1 and 2.
The momenta in the loop are
\beq \vec p_{12}=\vec{q}+\vec{k}_1 ,\qquad
\vec p_{23}= \vec{q}- \vec{k}_3
,\qquad
\vec p_{31}=\vec{q}.\eeq
Following~\cite{1805.02656,1907.10624}, we compute the diagram in the squeezed limit 
$ k_3 \ll k_{1,2} $.
The orientation of $\vec k_3$ does not affect the universal
oscillatory cosmo-collider signal~\cite{1805.02656,1907.10624},
so we assume that $\vec{k}_3$ is parallel to $\vec{k}_{1,2}$:
\beq \vec{k}_1 = (0,0,k_1), \qquad \vec k_2 = - \vec k_1 - \vec k_3 \simeq - \vec{k}_1,\qquad
\vec{k}_3\simeq (0,0,k_3).\eeq
As discussed below eq.\eq{Ek},
fermion pair production occurs predominantly 
when $k/a\approx \mu_A \pm m$, i.e.\ at $\eta \approx - \tilde{\mu}_A/k$~\cite{1805.02656,1907.10624,1910.12876}.
Thus, the long-wavelength $k_3$ mode leads to the earliest, dominant fermion production
at $\eta_3 \simeq -\tilde{\mu}_A/k_3$, and the two fermions are later absorbed at $\eta_{1,2}$,
by which time their energy has increased to $E\sim \mu_A$.
Approximating the hard momentum as
\beq \vec{p}_{12}\simeq \vec{k}_1\eeq
and the soft propagators at the hard vertices as described below,
the result becomes a sum of terms, each with a factorised form~\cite{2304.13295,2605.21419}
\beq \sum \bigg(\int d\eta_{1,2}\cdots\bigg) \bigg(\int d^3q \cdots\bigg) \bigg(\int d\eta_3\cdots\bigg).\eeq
Previous computations also approximated the soft 13 and 23 propagators by their late-time limit,
which allows the soft $\eta_3$ and $d^3q$ integrals to be performed analytically.
This approximation is summarised in appendix~\ref{app:late}, where we show that it 
underestimates the result.
We here keep the complete fermion modes at the soft vertex $\eta_3$.
On the other hand, at the hard vertices one has $|p_{23}\eta_2|,|p_{31}\eta_1|\sim k_3/k_1\ll 1$,
so that the soft fermion modes can be replaced by their late-time limit
\beq \label{eq:Wlate}
W_{\kappa,i\lambda}(x) \stackrel{|x|\to 0}\simeq \frac{\Gamma(-2i\lambda)}{\Gamma(\frac12-i\lambda-\kappa)}\,x^{\frac12+i\lambda} + (\lambda\to-\lambda).\eeq
The first term produces the oscillatory signal $(k_L/k_S)^{2i\lambda}$ in eq.\eq{shapefNL}; the second term produces its complex conjugate.
Furthermore, since the hard vertices occur later than the soft vertex, 
the soft propagators reduce to the Wightman functions of eq.\eq{DSK}; for the dominant $a_3=+1$,
\beq \begin{aligned} D_{a_2a_3}(\vec p_{23},\eta_2,\eta_3)&\simeq -i\sum_h u_{\vec p_{23},h}(\eta_2)\otimes\bar u_{\vec p_{23},h}(\eta_3),\cr
D_{a_3a_1}(\vec p_{31},\eta_3,\eta_1)&\simeq i\sum_{h'} v_{\vec p_{31},h'}(\eta_3)\otimes\bar v_{\vec p_{31},h'}(\eta_1).\end{aligned}\eeq
So the soft vertex enters the fermion trace through the spinor bilinear
$\bar u_{\vec p_{23},h}(\eta_3)\,\Gamma^{\mu_3} v_{\vec p_{31},h'}(\eta_3)$,
which describes the creation of a fermion pair at $\eta_3$ by the long-wavelength mode $\vec k_3$.

\smallskip

The integrals over $\eta_{1,2}\equiv -z_{1,2}/k_1$ can be expanded in terms of seed integrals
over the dimensionless $z_{1,2}$:
\beq \label{eq:SeedI}
\int_0^{\infty}\!\!\int_0^{\infty} {d}z_1\,{d}z_2\,
e^{-i(z_1+z_2)}
z_1^{p_1}
z_2^{p_2}
W_{w_1,\,i\mu}(-2iz_1)
W_{w_2,\,i\mu}(+2iz_2)
\theta(z_2-z_1) =
2{\cal I}^{p_1+1/2,p_2+1/2}_{w_1,w_2,\mu}.
\eeq
The results of~\cite{2605.21419} for the values of $w_{1,2} = \pm1/2 \pm i \tilde\mu_A$ that arise in the present computation
can be unified into
{\small\beq
\begin{aligned}
{\cal I}^{p_1p_2}_{w_1,w_2,\mu} &=
i \pi \frac{e^{-\frac{i\pi}{2}(p_1+p_2)}}{2^{2+p_1+p_2}}
\mathrm{csch}(2\pi\mu)\,
\Gamma(2+p_1+p_2)
\Bigg[
\frac{e^{-\pi\mu}\Gamma(1+p_1-i\mu)}
{\Gamma\!\left(\frac12-w_1+i\mu\right)}
\Gamma(2+p_1+p_2-2i\mu)
\\
&\times
{}_4\widetilde F_3
\!\left(
\begin{matrix}
2+p_1+p_2,\,
1+p_1-i\mu,\,
\frac12-w_1-i\mu,\,
2+p_1+p_2-2i\mu
\\[1mm]
2+p_1-i\mu,\,
\frac52+p_1+p_2-w_2-i\mu,\,
1-2i\mu
\end{matrix}
\,\bigg|\,-1
\right)
-\left(\mu\rightarrow-\mu\right)
\Bigg]
\end{aligned}
\eeq}
where $_4 \tilde{F}_3$ is the regularised hypergeometric function~\cite{2605.21419}.
In the soft integral, the $d^3q$ integral is conveniently performed in terms of the elliptic variables~\cite{2611.23220}
\beq z = \frac{k_3}{p_{23}+p_{31}},\qquad w = \frac{p_{31}-p_{23}}{k_3},\qquad
\int\frac{d^3q}{(2\pi)^3} = \frac{k_3^3}{32\pi^2}\int_0^1 \frac{dz}{z^4}\int_{-1}^{1}dw\,(1-w^2z^2),\eeq
such that $p_{23}=k_3(1-wz)/2z$ and $p_{31}=k_3(1+wz)/2z$.
The large-loop-momentum region corresponds to $z\to 0$, and
the infra-red regions where one soft momentum vanishes correspond to $z\to 1$ with $w\to\pm1$.
The $\eta_3$ integral is performed rotating the integration contour as $\eta_3 = i a_3 u z/k_3$, so that
$u = (p_{23}+p_{31})|\eta_3|$ is the time of the soft vertex in units of the energy of the produced fermion pair.
The external wave function becomes $e^{ia_3k_3\eta_3} = e^{-zu}$, 
and the fermion modes at the soft vertex become Whittaker functions of the real positive arguments 
$2p_{23}|\eta_3| = (1-wz)u$ and $2 p_{31}|\eta_3|=(1+wz)u$.
The soft integrals can thereby be expanded in terms of the soft seed integrals
\beq\label{eq:SeedS}
\mathcal{S}^{n}_{w_1,w_2,\lambda}(z,w)=\int_0^\infty du\, u^{n-1}e^{-zu}\,
W_{w_1,i\lambda}\big((1-wz)u\big)\,W_{w_2,i\lambda}\big((1+wz)u\big)
\eeq
where $n=0,1$ and $w_{1,2}=\pm\frac12\pm i\tilde\mu_A$.
At fixed $z,w$ each soft seed is a one-dimensional integral, regular at $u\to 0$ and exponentially damped at large $u$.
The hard-vertex factors of eq.\eq{Wlate} provide an overall $z^{-3-2i\lambda}$, 
so that the large-$q$ region is divergent.
As in dimensional regularisation, 
the divergence is a pure power and is removed by analytic continuation in the exponent:
writing the $w$-integrated integrand as $z^{-3-2i\lambda}F(z)$ with $F(z)=\sum_n F_n z^n$ regular at $z=0$, and taking the Finite Part:
\beq\label{eq:FP}
\hbox{FP}\int_0^1 dz\, z^{-3-2i\lambda} F(z) = \sum_{n=0}^{N}\frac{F_n\, z_0^{n-2-2i\lambda}}{n-2-2i\lambda}+
\int_{z_0}^1 dz\,z^{-3-2i\lambda}F(z)\eeq
for any $0<z_0<1$ and $N$ large enough.
This is the same prescription that dimensional regularisation applies to the loop integrals of eq.\eq{Bab}:
the discarded power divergences are analytic in $k_3$ and do not contribute to the oscillatory signal.

\smallskip

Terms with Schwinger--Keldysh indices $a_{1,2,3}=+1$ dominate.
Terms with $a_3=-1$ are exponentially suppressed in the large-$\tilde{\mu}_A$ limit,
and terms with $a_3=+1$ and one or both hard vertices on the negative branch are suppressed by powers of $1/\tilde{\mu}_A$.
The configuration $a_{1,2,3}=-1$ instead dominates the complex-conjugate contributions.
With these ingredients the computation proceeds similarly to
flat-space Feynman diagram computations: a software package expands eq.\eq{tau3} and  decomposes the terms into seed integrals.
All terms factorise into a common soft integral, times a combination of hard seeds.

\begin{figure}[t]
$$\includegraphics[width=0.7\textwidth]{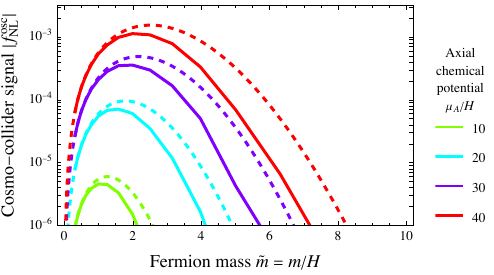}$$
\begin{center}
\caption{\em Amplitude of the oscillatory cosmo-collider signal.
The continuous curves show the full result.
The dashed curves show the asymptotic approximation of eq.\eq{fNLasy}.
Here $N_c=3$ and $\tilde{\mu}_V=0$.
The maximal chemical potential allowed by unitarity is $\mu_{A,V}\sim 60H$~\cite{1805.02656}.
\label{fig:ModularCosmofNL}}
\end{center}
\end{figure}

\subsection{Final result}
We find a shape function as in eq.\eq{shapefNL} with coefficient
\beq  \label{eq:fNLres}
f_{\rm NL}^{\rm osc} \approx \frac{i N_c P_\zeta\,\tilde\mu_A^2\,\Gamma(-2i\lambda)^2}{16\,(\tilde\mu_A+\lambda)}\,e^{-\pi(\tilde\mu_A+\lambda)}\,
c_{\rm hard}\, c_{\rm soft}.\eeq
Keeping only the dominant terms for large $\lambda$, the hard propagators give
\beq\begin{aligned}
c_{\rm hard}= &
\frac{e^{\pi  \tilde{\mu }_A} (\tilde{\mu }_A^2+\tilde{\mu }_V^2)}{ \tilde{\mu }_A}
 \mathcal{I}_{\frac{1}{2}-i \tilde{\mu }_A,\frac{1}{2}+i \tilde{\mu }_A,\lambda }^{i \lambda ,1+i \lambda } +\\
 &
 +
 (\tilde{\mu }_A^2-\tilde{\mu }_V^2)   \frac{ \tilde{\mu }_A+\lambda }{ \tilde{\mu }_A}
 \bigg[
 \frac{e^{\pi  \tilde{\mu }_A}}{2}
  \mathcal{I}_{\frac{1}{2}-i \tilde{\mu }_A,-\frac{1}{2}+i \tilde{\mu }_A,\lambda }^{i \lambda ,i \lambda }
  +
2e^{-\pi  \tilde{\mu }_A}   
   \mathcal{I}_{-\frac{1}{2}+i \tilde{\mu }_A,\frac{1}{2}-i \tilde{\mu }_A,\lambda }^{1+i \lambda ,1+i \lambda }\bigg].
   \end{aligned}
\eeq
This can be approximated as $ |c_{\rm hard}| \stackrel{\lambda\to\infty}\simeq 2\sqrt{\pi}\,\lambda ^{7/2}$.
The soft vertex and the loop integral give
\beq\label{eq:Bsoft}
c_{\rm soft} = \hbox{FP}\int_0^1 \frac{dz}{z^{3+2i\lambda}}\int_{-1}^1 \frac{dw}{(1-w^2z^2)^{\frac12-i\lambda}} \sum_{j=1}^8 c_j
\big[A_j\, \mathcal{S}^{0}_{w_j,w'_j,\lambda}+B_j\, \mathcal{S}^{1}_{w_j,w'_j,\lambda}\big]\eeq
where the eight terms are listed in table~\ref{tab:channels}.
Appendix~\ref{app:seeds} outlines how soft integrals are evaluated.
Combining channels related by $w\to-w$, table~\ref{tab:channels} reduces to the four channels found in~\cite{2611.23220} for a Majorana fermion,
with identical kernels and proportional weights.
Our $f_{\rm NL}^{\rm osc}$ agrees with their result, being twice it as appropriate for a Dirac fermion,
once the different normalisations of $f_{\rm NL}$ are taken into account.
%
\begin{table}[t]
\begin{center}
\renewcommand{\arraystretch}{1.4}
$\begin{array}{c|cc|c|c|c}
j & w_j & w'_j & c_j & A_j & B_j\\ \hline
1 & -\frac12-i\tilde\mu_A & \frac12-i\tilde\mu_A & -\tilde m^2 e^{3\pi\tilde\mu_A}/g_1^2 & w\, P_- & +z(1+w)P_-\\
2 & \frac12-i\tilde\mu_A & -\frac12-i\tilde\mu_A & +\tilde m^2 e^{3\pi\tilde\mu_A}/g_1^2 & w\, P_- & -z(1-w)P_-\\
3 & -\frac12+i\tilde\mu_A & \frac12+i\tilde\mu_A & -\tilde m^4 e^{-\pi\tilde\mu_A}/g_2^2 & w\, P_+ & +z(1+w)P_+\\
4 & \frac12+i\tilde\mu_A & -\frac12+i\tilde\mu_A & +\tilde m^4 e^{-\pi\tilde\mu_A}/g_2^2 & w\, P_+ & -z(1-w)P_+\\
5 & -\frac12-i\tilde\mu_A & -\frac12+i\tilde\mu_A & +\tilde m^4 e^{\pi\tilde\mu_A}/g_1g_2 & (1-w^2)Q_+ & (1-w^2)z(1-z)Q_+\\
6 & -\frac12+i\tilde\mu_A & -\frac12-i\tilde\mu_A & -\tilde m^4 e^{\pi\tilde\mu_A}/g_1g_2 & (1-w^2)Q_- & (1-w^2)z(1-z)Q_-\\
7 & \frac12-i\tilde\mu_A & \frac12+i\tilde\mu_A & +\tilde m^2 e^{\pi\tilde\mu_A}/g_1g_2 & (1-w^2)Q_+ & (1-w^2)z(1+z)Q_+\\
8 & \frac12+i\tilde\mu_A & \frac12-i\tilde\mu_A & -\tilde m^2 e^{\pi\tilde\mu_A}/g_1g_2 & (1-w^2)Q_- & (1-w^2)z(1+z)Q_-
\end{array}$
\caption{\em\label{tab:channels} The eight terms contributing to the soft factor $c_{\rm soft}$ of eq.\eq{Bsoft}.
The Whittaker index $w_j$ ($w'_j$) refers to the fermion with momentum $p_{23}$ ($p_{31}$), with argument $(1-wz)u$ ($(1+wz)u$).
Here $g_1 = \Gamma(-i(\tilde\mu_A+\lambda))$, $g_2=\Gamma(1+i\tilde\mu_A-i\lambda)$,
$P_-=(z^2-1)[1+w-e^{2\pi(\lambda-2\tilde\mu_A)}(1-w)]$, $P_+=(z^2-1)[w-1+e^{2\pi(\lambda+2\tilde\mu_A)}(1+w)]$ and
$Q_\pm = e^{2\pi\lambda}(1\pm z)+(1\mp z)$.}
\end{center}
\end{table}
In the limit of large $\tilde\mu_{A}$ one gets the asymptotic expression 
\beq \label{eq:fNLasy}
|f_{\rm NL}^{\rm osc}| \stackrel{\tilde{\mu}_{A}\to\infty}\simeq  
N_c   \tilde{m}^2 P_{\zeta } \frac{\pi ^{5/2}  \tilde{\mu }_A^{3}  }{8  \sqrt{2}}
e^{-2\pi \tilde m^2/\tilde{\mu}_A}
\eeq
valid for $\tilde m^2\lesssim \tilde\mu_A/5$.
Our result is illustrated in fig.\fig{ModularCosmofNL}. 
$f_{\rm NL}^{\rm osc}$ is too small to be observed in near-future experiments:
eq.\eq{fNLasy} is maximal for $\tilde m^2 \approx \tilde\mu_A/2\pi$, where $f_{\rm NL}^{\rm osc}\approx 6~10^{-10} \tilde\mu_A^4$,
giving $f_{\rm NL}^{\rm osc}\lesssim 10^{-2}$ up to the unitarity limit $\tilde\mu_A\approx 60$.
A larger $f_{\rm NL}^{\rm osc}$ can arise by abandoning the minimal scenario where primordial inhomogeneities are generated by
inflaton fluctuations and considering the curvaton or modulated reheating~\cite{hep-ph/0109214,astro-ph/0303591}.
Barring such possibilities, the result is small because eq.\eq{fNLasy} is suppressed by about $1/\tilde\mu_A^2 \tilde m$
compared to the estimate of eq.\eq{fNLapprox} claimed in previous analyses~\cite{1805.02656,1908.00019,1907.10624,1910.12876,2010.04727}.
The differences are the following:
\begin{itemize}
\item
We use 4-component Dirac fermions $\Psi$, so that only $\Psi\bar\Psi$ propagators appear. 
In contrast, previous analyses employed 2-component Weyl spinors $\psi$, which require a combination of $\psi\bar\psi$ and $\psi\psi$ propagators. 
Our Dirac trace automatically incorporates all such contributions in a unified way.
One difference is that the anticommutation relation eq.\eq{anticom} implies that modes with helicity $h=+1$ do not propagate into modes with helicity $h=-1$.
Contributions corresponding to helicity-mixing propagation should be 
omitted from previous computations.
This mildly suppresses the final powers of $\tilde\mu_A$ and $\tilde m$ compared to eq.\eq{fNLapprox}.

\item In~\cite{1805.02656} the hard 12 propagator was approximated by its value at fixed $\eta_1 =\eta_2 =- \tilde{m}/2k_1$,
so that the $\eta_{1,2}$ integrals also acquire the simpler form of eq.\eq{tauintegral}.
Following~\cite{2605.21419}, we performed the integral in terms of the seed function in eq.\eq{SeedI}.
This mildly reduces the value of $f_{\rm NL}^{\rm osc}$.

\item In~\cite{1805.02656,1907.10624} the soft loop integral was approximated as
\beq \int \frac{d^3q}{(2\pi)^3} \approx \frac{k_3^3}{(2\pi)^2}.\eeq
This approximation turns it into a phase-space integral that reconstructs the fermion density enhanced by chemical potentials.
The loop integral instead shows that different momentum modes partially interfere at the amplitude level rather than summing to their total abundance.

\item We included the vector chemical potential.
Although the modular theory considered here is different, it provides chemical potentials similar to those in previous works.
\end{itemize}
Given that the final result is small, additional suppressed contributions might be relevant.
We neglected the `local' contribution to $f_{\rm NL}$, which arises from virtual fermions mediating a loop correction to cubic $\tau$ interactions.
More generally, single-field consistency relations require small inflaton cubic interactions,
implying $f_{\rm NL}^{\rm local} \sim 10^{-2}$ in slow-roll models~\cite{Maldacena:2002vr,Creminelli:2004yq}.

\subsection{Effects of anomalous couplings to vectors}
We now consider the one-loop-suppressed anomalous couplings of the modulus $\tau$ to the SM vectors, eq.\eq{Lanom}.
After integration by parts, these couplings give an interaction of $\partial_\mu\tau$ with the Chern–Simons current.
So it induces a chemical potential-like term~\cite{1110.3327}
\begin{equation}
\mu_{i} \;\sim\; \frac{g_{i}^2}{(4\pi)^2}\,\frac{\theta_{i}'(\tau)\,\dot\tau_0}{f} \sim 10^{-3}\mu_{A},\qquad i = \{1,2,3\}.
\end{equation}
More precisely, it affects the dispersion relations of transverse vectors as $E_\pm^2 = k^2 \pm 2k\mu_i$.
As a result, despite being loop-suppressed, this term triggers a tachionic instability for one vector helicity in a range of $k$.
This leads to vector particle production possibly enhanced by $e^{\pi \tilde\mu_{i}}$, where
$\tilde\mu_{i}\equiv \mu_{i}/H$,
up to backreaction and other issues~\cite{1806.08769,2309.04254}.
In our theory, the combinations $\tilde\mu_{i}$ typically remain below unity, even when the fermion chemical potentials
approach the maximal value allowed by unitarity, $\tilde\mu_{A,V} \sim 60$.
In this regime gauge-field production is inefficient and does not compete with the fermionic effects discussed above.
A possible exception arises if some modular function $\theta'_{i}$ varies rapidly along the inflationary trajectory.
Barring this possibility, we expect that the anomalous couplings to vectors do not give a dominant
contribution to the cosmological collider signal.

\section{Conclusions}\label{sec:concl}
We studied cosmological collider signals in string-motivated modular-invariant extensions of the Standard Model, 
where CP violation arises dynamically from the vacuum expectation value of a scalar modulus $\tau$.
If it acts as the inflaton, the time-dependent phases of the SM Yukawa couplings induce effective chemical potentials for SM fermions and for the Higgs.
As a result, the Higgs develops an inflationary condensate,
giving SM fermions a large inflationary mass $m$.

Theories with a similar structure --- where chemical potentials are associated with a broken 
U(1) --- have previously been introduced in an ad hoc manner to enhance cosmological collider signals~\cite{1805.02656,1908.00019,1907.10624,1910.12876,2010.04727,2507.22978}.
Modular-invariant theories offer a concrete and predictive framework in which analogous physics emerges naturally, 
through time-dependent Yukawa phases and the resulting effective chemical potentials.
Notably, this setup requires only a single additional degree of freedom, the modulus $\tau$, 
making it a minimal and well-motivated extension of the Standard Model.

In these models, SM fermions mediate a one-loop contribution to the inflaton bispectrum of oscillatory type in the squeezed limit.
Previous estimates~\cite{1805.02656,1908.00019,1910.12876} suggested that this signal could be strongly enhanced by fermion chemical potentials. In particular, an observable cosmological-collider signal was expected if a Standard Model fermion, 
such as the bottom or top quark, has a mass $m\sim H$ and an axial chemical potential $\mu_A\sim 10H$.

We carefully quantised four-component Dirac fermions with generic axial and vector chemical potentials in de Sitter. 
This allows for a unified treatment of the propagators and makes the helicity structure manifest. 
In particular, modes with definite helicity propagate independently, and no helicity-mixing contributions arise. 
This leads to a difference from previous results obtained using 2-component formalisms.
More importantly, we evaluated, rather than estimated, the loop integral, finding that it
does not simply reproduce the chemical-potential-enhanced fermion density. 
Contributions from different loop momenta partially interfere, suppressing the signal.
This was first noticed in~\cite{2605.21419,2611.23220}.
The resulting signal, eq.\eq{fNLasy}, remains below the sensitivity of near-future experiments.

\smallskip


We also considered anomalous couplings to gauge fields. 
Despite the possibility of tachionic production, loop suppression makes gauge-field effects subdominant unless modular functions vary rapidly.

 \small \paragraph{Acknowledgements.}  We thank Xingang Chen, Anish Ghoshal,  Chengcheng Han, Hong-Jian He,
 Sang Hui Im, Davide Racco, Zhehan Qin, Linghao Song, Zhong-Zhi Xianyu, Zhaohui Xu, Jingtao You, and Yuhang Zhu for discussions. S.A. is supported by the Japan Science and Technology Agency (JST) 
as part of Adopting Sustainable Partnerships for Innovative Research Ecosystem (ASPIRE), 
grant JPMJAP2318, and by JSPS KAKENHI Grant Number 26K07078. 
The computation of soft integrals beyond the late-time approximation was performed by Claude.
AS thanks Anthropic for a `Claude Team plan for scientists' discount.

\appendix

\section{Computing the soft integrals}\label{app:seeds}
The difference between the full computation and the late-time approximation is contained in the soft seeds of eq.\eq{SeedS}.
Their structure becomes transparent writing each Whittaker function as a superposition of exponentials,
\beq\label{eq:Wlaplace} W_{\kappa,i\lambda}(x) = x^{\frac12-i\lambda}\int_0^\infty dt\, e^{-x(\frac12+t)}\, [g_{\kappa0}(t)+x\, g_{\kappa1}(t)],\qquad
\begin{aligned}
g_{\kappa0}(t)&=-\beta\, t^\alpha(1+t)^{\beta-1}/\Gamma(1+\alpha),\cr
g_{\kappa1}(t)&=t^\alpha(1+t)^\beta/\Gamma(1+\alpha),\end{aligned}\eeq
where $\alpha=\frac12-i\lambda-\kappa$ and $\beta=-\frac12-i\lambda+\kappa$.
This is the standard integral representation of $W$ after one integration by parts in $t$, valid for $\Re\alpha>-1$.
It decomposes each fermion mode into flat-space-like waves with energies $(1+2t)$ times the momentum.
The region of large $t$, where $g_{\kappa a}(t)$ reduce to power laws, reproduces the late-time behaviour $x^{\frac12\pm i\lambda}$ of eq.\eq{Wlate};
the region $t\sim 1$ describes the mode around the time when it is produced.
Inserting eq.\eq{Wlaplace} into eq.\eq{SeedS}, the time integral becomes elementary, as in flat-space perturbation theory:
\beq\label{eq:SeedSrep}
\mathcal{S}^{n}_{w_1,w_2,\lambda}=(1-w^2z^2)^{\frac12-i\lambda}
\sum_{a,b=0,1}(1-wz)^a(1+wz)^b\,\Gamma(n+1+a+b-2i\lambda)\int_0^\infty\!\!\! dt_1\int_0^\infty \!\!\!dt_2\,
\frac{g_{w_1a}(t_1)\,g_{w_2b}(t_2)}{E^{n+1+a+b-2i\lambda}}\eeq
where $E=1+t_1+t_2+z\,[1+w(t_2-t_1)]$ is the total energy at the soft vertex in units of $p_{23}+p_{31}$.
Inserting this into eq.\eq{Bsoft}, the factors $(1-w^2z^2)^{\frac12-i\lambda}$ cancel.
Since $E$ is linear in $w$ and, after writing $z=1/y$, $z^{-3-2i\lambda}E^{-(n+1+a+b-2i\lambda)}$ becomes an integer power of $y$ times a power of a function linear in $y$,
the $w$ integral and the finite-part $z$ integral can be performed in closed form for each channel.
The soft factor $c_{\rm soft}$ thereby reduces to a two-dimensional integral of elementary functions over $t_1,t_2$.
Performing also these last integrals gives the Meijer $G$ functions of~\cite{2611.23220}.

\subsection{Late-time approximation}\label{app:late}
\label{sec:latelime}
Previous computations approximated the loop diagram using the late-time limit $\eta\to 0$ of all fermion propagators.
This corresponds to replacing each Whittaker function in eq.\eq{SeedS} by its small-argument limit, eq.\eq{Wlate};
the $u$ integral then gives Gamma functions, and the $d^3q$ integral gives the $B$ functions of eq.\eq{Bab}.
The late-time approximation is simpler, but underestimates the final result.
It  is not accurate because the soft fermion pair is produced at
$|\eta_3|\sim \tilde\mu_A/k_3$, where $|p\,\eta_3|$ is not small for the loop momenta that dominate the integral.
Keeping the complete modes at the soft vertex, as in section~\ref{sec:detailed}, avoids this approximation.

More in detail, in the late-time limit the components of all $u_\pm, v_\pm$ modes acquire the form 
\beq (-2k\eta)^{-i\tilde{\mu}_V} [\ldots (i k\eta)^{i \lambda} + \ldots (i k\eta)^{-i \lambda }]\eeq
where $\ldots$ denotes functions of $\tilde m,\tilde{\mu}_A$.
So the products of spinors at times $\eta_1$ and $\eta_2$ that appear in propagators simplify to
\beq
(\eta_1/\eta_2)^{i \tilde{\mu}_V}[\ldots (\eta_1 \eta_2)^{-i\lambda} + \ldots (\eta_1/ \eta_2)^{i\lambda}+ \hbox{h.c.}].\eeq
We neglect higher-order $1+ {\cal O}(k\eta)$ terms, because the corrections are predominantly imaginary and affect the oscillatory behaviour.
We discard the `local' terms $(\eta_1/ \eta_2)^{i\lambda}$, since they are not associated with particle production: the enhancement from chemical potentials arises from the non-local part of the propagators, which encodes on-shell propagation and is proportional to the fermion density, while purely virtual (local) contributions do not receive such enhancement. 
Then the remaining non-local terms simplify, and all $D_{\pm\pm}$ propagators have a common limit.
Writing only the contribution from helicity $h=1$, which is enhanced by a chemical potential ${\mu}_A>0$, one has
\beq \label{eq:D+-D-}
{D}_{+-}^{\rm non~local} \simeq {D}_{-+} ^{\rm non~local}\simeq  - i c_+(\lambda)
 \left(\frac{\eta_1}{\eta_2}\right)^{- i \tilde{\mu}_V}(4k^2\eta_1\eta_2)^{+i \lambda}    \Pi  \Pi_+ + (\lambda\to - \lambda).
\eeq
Here $\Pi$ and $\Pi_h$ are projectors
\beq \label{eq:projectors}
\Pi = \frac{\tilde{\mu}_A -\lambda \gamma^5 -\tilde{m}\gamma^0\gamma^5 }{2\tilde{\mu}_A},
\qquad
\Pi_h(\hat{k})=\frac{1 + h\,\gamma^5  \gamma^0 \vec\gamma \cdot \hat{k}}{2}\eeq
and
\beq c_+ (\lambda)= \frac{2\tilde{\mu}_A e^{\pi  \tilde{\mu} _A} \Gamma (-2 i \lambda)^2}{\tilde{m} 
\Gamma (-i ( \tilde{\mu } _A+\lambda ))
   \Gamma (i (\tilde{\mu }_A-\lambda))}.\eeq
It reduces to $|c_+| \simeq\tilde m  \sqrt{\pi/\tilde{\mu}_A}$ for large $\tilde{\mu}_A$ and small $\tilde{m}$. 
Similar expressions hold for the helicity $h=-1$, and roughly correspond to flipping $\tilde{\mu}_A\to -\tilde{\mu}_A$.
As a result, $|c_{-}(\lambda)/c_{+}(\lambda)|\simeq e^{-2\pi \tilde{\mu}_A}$.

\medskip

In the loop diagram of fig.\fig{Feyn2} the 13 and 23 propagators are approximated by their late-time, non-local limit,
retaining only the helicity that is enhanced by the chemical potentials, $h=+1$.
Expanding the product of such 13 and 23 propagators
leads to some terms in which powers of $\eta_3$ cancel.
We discard such  local terms,
in addition to the other local terms already discarded by approximating $D_{13,23}$ with 
their non-local counterparts $D_{13,23}^{\rm non~local}$.
In the hard propagator $D_{12}$ both helicities $h=\pm 1$ must be included, 
and the various components $D_{a_1 a_2}$ are inequivalent.
The soft integrals can then be performed analytically as follows:
\begin{itemize}
\item The integrals over $\eta_3$ have the form
\begin{align}\label{eq:tauintegral}
\int_{-\infty}^0 d \eta \, e^{  ia  k\eta} \, (-\eta)^{p-1}= \frac{ \Gamma(p)  }{(iak)^{p}}
\end{align}
for various values of the complex powers $p$. 

\item The loop integrals over $d^3q$ can be expanded in terms of the massless Passarino-Veltman $B$ function in $d=3$ dimensions
\beq\label{eq:Bab}
\int\frac{d^dq}{(2\pi)^3}  \frac{1}{(\vec q)^{2a} (\vec{q}+\vec{k})^{ 2b}}
=\frac{B_{ab} }{(4\pi)^{d/2} (k^2)^{a+b-d/2}} ,\qquad
B_{ab}= 
\frac{ 
\Gamma\!(\frac{d}{2}-b)
\Gamma\!(\frac{d}{2}-a)
\Gamma\!(a+b-\frac{d}{2})
}
{\Gamma(a)\Gamma(b)\Gamma(d-a-b)} .
\eeq
Scalar products appearing in the numerators
are reduced to this form using $\vec{q}\cdot\vec k =[ (\vec{q}+\vec{k})^2 - q^2 - k^2]/2$.
\end{itemize}
The result is
\beq f_{\rm NL}^{\rm osc}\big|_{\rm late} \approx \sqrt{\pi } N_c  P_{\zeta }  2^{-5+4 i \lambda } \tilde{m}^2 \Gamma (2+2 i \lambda) 
e^{\pi  \lambda } \, c_{\rm hard}\, c_+^2 B\eeq
with the same $c_{\rm hard}$ as in eq.\eq{fNLres}, and
\beq \label{eq:bubbleres}
B = 2 B_{-\frac{1}{2}+i \lambda ,\frac{1}{2}+i \lambda }-B_{\frac{1}{2}+i \lambda ,\frac{1}{2}+i \lambda }+2 B_{i \lambda ,i\lambda },\qquad
|B|\stackrel{\lambda\to\infty}\simeq
\frac{\lambda^{-5/2}}{8\sqrt{2}}
\eeq
which is more strongly suppressed at large $\lambda$ than the individual terms.
For large imaginary values of $a=b$ one has
$ | B_{aa}| \simeq {2^{4\Re a - 9/2}}/{ (\Im a)^{3/2}}$,
which means that different momentum modes interfere at the amplitude level rather than summing to their total abundance~\cite{2605.21419}.
Including the prefactor, $|c_+^2 B|\approx \pi \tilde m^2/8\sqrt2\tilde\mu_A\lambda^{5/2}$.
In the limit of large $\tilde\mu_{A}$ the late-time approximation gives
\beq \label{eq:fNLasylate}
|f_{\rm NL}^{\rm osc}|_{\rm late} \stackrel{\tilde{\mu}_{A}\to\infty}\simeq  
N_c   \tilde{m}^4 P_{\zeta } \frac{\pi ^{5/2}  \tilde{\mu }_A^{3/2}  }{32  \sqrt{2}}
e^{-2\pi \tilde m^2/\tilde{\mu}_A}
\eeq
Keeping the complete fermion modes at the soft vertex removes most of this destructive interference.
For $\tilde m^2\ll\tilde\mu_A$ the full result of eq.\eq{fNLasy} gives
$ |\sfrac{f_{\rm NL}^{\rm osc}}{f_{\rm NL}^{\rm osc}{}_{\rm late}} |\simeq \sfrac{4\tilde\mu_A^{3/2}}{\tilde m^2}$,
in agreement with~\cite{2611.23220,2611.31214}.

\end{document}